\documentclass[aps,prd,preprint,nofootinbib,a4paper,11pt,superscriptaddress]{revtex4-1}

\usepackage[T2A,T1]{fontenc}
\usepackage[centertags]{amsmath}
\usepackage{amsfonts}
\usepackage{amssymb}
\usepackage[sort&compress]{natbib}
\usepackage[hypertex]{hyperref}

\DeclareMathOperator{\pr}{pr}
\DeclareMathOperator{\re}{Re}
\DeclareMathOperator{\im}{Im}

\newcommand{\lan}{\langle}
\newcommand{\ran}{\rangle}

\newcommand{\e}{\varepsilon}
\newcommand{\vf}{\varphi}

\newcommand{\al}{\alpha}
\newcommand{\be}{\beta}
\newcommand{\ga}{\gamma}

\newcommand{\de}{\delta}

\newcommand{\la}{\lambda}
\newcommand{\La}{\Lambda}
\newcommand{\ups}{\upsilon}

\begin{document}


\title{Asymptotics of physical solutions to the Lorentz-Dirac equation for a planar motion in constant electromagnetic fields}

\date{\today}

\author{P.O. Kazinski}
\email[E-mail:]{kpo@phys.tsu.ru}
\affiliation{Physics Faculty, Tomsk State University, Tomsk, 634050 Russia}
\affiliation{Institute of Monitoring of Climatic and Ecological Systems, SB RAS, Tomsk, 634055 Russia}
\author{M.A. Shipulya}
\email[E-mail:]{sma@phys.tsu.ru}
\affiliation{Physics Faculty, Tomsk State University, Tomsk, 634050 Russia}

\begin{abstract}

We present a study of planar physical solutions to the Lorentz-Dirac equation in a constant electromagnetic field. In this case, we reduced the Lorentz-Dirac equation to the one second order differential equation. We obtained the asymptotics of physical solutions to this equation at large proper times. It turns out that, in the crossed constant uniform electromagnetic field with vanishing invariants, a charged particle goes to a universal regime at large times. We found the ratio of momentum components which tends to a constant determined only by the external field. This effect is essentially due to a radiation reaction. There is not such an effect for the Lorentz equation in this field.

\end{abstract}


\maketitle

\section{Introduction}

The Lorentz-Dirac (LD) equation has an ill fame to suffer from the various type inconsistencies. The latter come from the higher derivative Schott term entering the LD equation and appear as the blowing up (runaway) and acausal solutions. However, despite of its undesirable features, we have to accept the LD equation as a correct one in its range of applicability by the following reasons. First, as it was shown in the seminal paper by Dirac \cite{Dir} and then more elaborately in \cite{Teit}, the LD equation stems from the energy-momentum conservation law provided a charged particle is sufficiently small and possesses neglible higher multipoles of a charge distribution. Second, under these assumptions the LD equation is a minimal evolutionary equation describing a radiation reaction which complies with all the symmetries of the model: the Poincare and reparameterization invariance. Furthermore, having made certain approximations, the LD equation was derived in the context of quantum electrodynamics (see, e.g., \cite{JoHu}), where the LD equation can be considered as a leading quasiclassical asymptotics to the Schwinger-Dyson equations for an electron. Therefore, the LD equation makes a physical sense and, under certain conditions, its solutions should give rise to predictions which can be observed in experiments. It is clear that the LD equation is valid in the range of energies and field strengths where the quantum corrections are neglible in comparison with the classical contribution. Rough general estimates of this range can be found, e.g., in \cite{Shen,Klepik}, and the more accurate analysis for the particular case of a constant homogeneous magnetic field is presented in \cite{Bagrov}. Various generalizations of the LD equation to include spin and higher multipoles \cite{Frenk}, or an interaction with non-Abelian gauge fields \cite{Wong} and gravity \cite{DeWBr}, to higher dimensions \cite{Kos}, to dyonic \cite{Rohrl} and massless charged particles \cite{rrmlp} are also known. All of them have the higher derivative terms and, hence, possess the same unwanted properties as the LD equation.

There is a coherent approach \cite{Bhabha,Plass,GuptaEl,Barut,Klepik,RohrlBook} how to extract a physical information from the LD equation and its analogues. It is based on the notion of a physical solution. In a general setting, it looks as follows. Given a system of interacting fields $\phi^\ga_a$ with the action functional
\begin{equation}\label{action0}
    S[\phi^\ga_1,\ldots,\phi^\ga_N]=\sum_{a=1}^N\la^{-1}_a S^0_a[\phi_a^\ga]+S_{int}[\phi^\ga_1,\ldots,\phi^\ga_N],
\end{equation}
where $\ga$ is a condensed index representing group and spacetime indices and spacetime points, $a$ numerates the fields, and $\la_a$ are some constants. Then the solution $\phi^\ga_a(\la)$ of the coupled system of equations of motion corresponding to the action \eqref{action0} at given initial and boundary conditions is called physical if there exist finite limits
\begin{equation}
    \lim_{\la_b\rightarrow0}\phi^\ga_a(\la),\qquad a=\overline{1,N},
\end{equation}
other lambdas being fixed. This regularity condition completely rules out the runaway solutions to the LD equation. Besides, its physical solutions are unambiguously specified by the six initial data -- three position coordinates and three momentum components -- as it should be in the realm of Newtonian mechanics.

Since the LD equation is nonlinear, it is hard to solve it even in simple external field configurations. Almost all the exact solutions to the LD equation can be found in \cite{Plass,RohrlBook,SokTer,DeRaStTr,GuptaEl}. In this paper, we address the problem of a finding and description of exact physical solutions to the LD equation in a constant homogeneous electromagnetic field. Moreover, we restrict ourself by a planar motion only. Even in this rather simple situation we did not succeed in finding the exact essentially planar (i.e. non linear) solutions to the LD equation. However, we reduce the LD equation to the one second order differential equation and investigate the asymptotics of its physical solutions at large times. For the constant homogeneous magnetic field this asymptotics was found in \cite{Plass}. As far as the constant electric and crossed fields are concerned these asymptotics, to our knowledge, are obtained for the first time. It turns out that, in the crossed field configuration, the LD equation possesses an attractor and the system passes into a universal regime at large times. After the lapse of time, the identical charged particles moving on the plane in such an electromagnetic field ``forget'' their initial data. Their trajectories become parallel and a certain ratio of momenta components tends to a constant which is independent of the initial conditions and determined by the external field only. This effect is essentially due to the radiation reaction. It is absent for the solution to the Lorentz equation in this field and can serve as an explicit manifestation of a validity of the classical radiation reaction theory in the domain of its applicability.

The paper is organized as follows. In Sec. \ref{general}, we present general formulas regarding a radiation reaction and define the physical solutions to the LD equation. Here we also give an integro-differential equation for the physical solutions. Sec. \ref{linplan} is a main part of the article. In Sec. \ref{linmot}, we briefly describe the linear motion and exact solutions to the LD equation in this case. In Sec. \ref{planmotgen}, we investigate the symmetries of the LD equation and provide the necessary and sufficient condition for the motion of a charged particle to be planar. In Sec. \ref{planmotsecor}, we derive the second order differential equation describing planar solutions to the LD equation. Sec. \ref{planmotasym} is devoted to the asymptotics of the physical solutions to the LD equation at large times. In Sec. \ref{planmotLL}, we consider the same problem in the framework of the so-called Landau-Lifshitz equation \cite{LandLifsh}. In conclusion, we summarize the main results of the paper and discuss the prospects for a further research.

\section{General formulas}\label{general}

Consider a particle with the charge $e$ and mass $m$ interacting with the electromagnetic field $A_\mu$ on Minkowski background $\mathbb{R}^{1,3}$ with the metric $\eta_{\mu\nu}=diag(1,-1,-1,-1)$ and coordinates $x^\mu$, $\mu=\overline{0,3}$. The action functional for such a system has the form
\begin{equation}\label{action particl}
    S[x(\tau),A(x)]=-m\int{d\tau\sqrt{\dot{x}^2}}-e\int{d\tau A_\mu
    \dot{x}^\mu}-\frac1{16\pi}\int{d^4xF_{\mu\nu}F^{\mu\nu}},
\end{equation}
where $x^\mu(\tau)$ defines the particle worldline, $F_{\mu\nu}:=\partial_{[\mu}A_{\nu]}$ is the strength tensor of the electromagnetic field (the square brackets denote an antisymmetrization without $1/2$)
\begin{equation}\label{fmunu}
    F_{\mu\nu}=\begin{bmatrix}
                 0 & E_x & E_y & E_z \\
                 -E_x & 0 & -H_z & H_y \\
                 -E_y & H_z & 0 & -H_x \\
                 -E_z & -H_y & H_x & 0 \\
               \end{bmatrix},
\end{equation}
and we take the system of units in which the speed of light $c=1$. In the proper time parameterization $\dot{x}^2=1$, the LD equation \cite{Lor,Dir} reads as
\begin{equation}\label{lde_ini}
    m\ddot{x}_\mu=eF_{\mu\nu}\dot{x}^\nu+\frac23e^2(\dddot{x}_\mu+\ddot{x}^2\dot{x}_\mu),
\end{equation}
where $F_{\mu\nu}$ is the strength tensor of the external electromagnetic field. Introducing the dimensionless quantities
\begin{equation}
    x^\mu\rightarrow m^{-1}x^\mu,\qquad\tau\rightarrow m^{-1}\tau,\qquad F_{\mu\nu}\rightarrow m^2e^{-1}F_{\mu\nu},
\end{equation}
we rewrite it in the form
\begin{equation}\label{lde}
    \dot{\ups}_\mu=f_\mu+\la(\ddot{\ups}_\mu+\dot{\ups}^2\ups_\mu),\qquad f_\mu:=F_{\mu\nu}\ups^\nu,
\end{equation}
where $\la:=2e^2/3$ and $m\ups^\mu:=m\dot{x}^\mu$ is the $4$-momentum of the particle.

The LD equation possesses unphysical solutions. Following \cite{Bhabha,GuptaEl}, we shall call the solution $x^\mu(\la,\tau)$ physical if it tends to the solution $x^\mu(0,\tau)$ of the corresponding Lorentz equation as $\la$ goes to zero. It is a realization of the general definition given in Introduction in the case of classical electrodynamics. According to this definition, the physical solution should be regular at the large mass $m$ and small $e^2$. From Eq. \eqref{lde_ini} we see that this requirement leads to a regularity of the physical solution with respect to the external field and the parameter $\la$ entering \eqref{lde}. The former simply follows from the general theorems regarding dependence of solutions to ordinary differential equations on a parameter (see, e.g., \cite{Golubev}), while the latter condition is not trivial. Also notice that all the known physically reasonable solutions to the LD equation are physical in the sense adopted by us. Some extra arguments in favor of this definition of physical solutions are given in Appendix.

We can find these solutions perturbatively as a (formal) series in $\la$. This perturbative scheme reduces the order of the LD equation and provides a unique solution to it at some fixed initial position and velocity of the particle. The first iteration of this perturbative procedure yields the Landau-Lifshitz equation \cite{LandLifsh}. It is not difficult to write an integro-differential equation which describes the physical solutions to the LD equation \cite{Barut,Plass,Klepik,RohrlBook}. In the proper time parameterization, it looks like
\begin{equation}\label{lde_integral}
    \dot{\ups}_\mu(\tau)=\pr_\mu^\nu(\tau)\int_0^\infty dte^{-t}\mathcal{P}_\nu(\tau+\la t),\qquad\pr_\mu^\nu:=\de^\nu_\mu-\ups_\mu\ups^\nu,
\end{equation}
where $\mathcal{P}_\mu=f_\mu+\la\dot{\ups}^2\ups_\mu$. Solutions to Eq. \eqref{lde_integral} are solutions to the LD equation \eqref{lde} with $\ups_\mu\dot{\ups}^\mu=0$. It is the latter requirement that gives rise to the projector entering Eq. \eqref{lde_integral}.  The solutions of this equation tend to the solutions of the Lorentz equation at $\la\rightarrow0$. Expanding Eq. \eqref{lde_integral} in a series in $\la$, we see that solutions to Eq. \eqref{lde_integral} are those solutions to the LD equation which are obtained from it by the aforementioned perturbative scheme. If we knew all the terms of the perturbation series for the acceleration $\dot{\ups}(\tau,\la)$ then formula \eqref{lde_integral} would tell us that this series in $\la$ must be summed by the Borel method \cite{Hardy}. So, if the following conditions are satisfied at some fixed initial position and velocity
\begin{enumerate}
  \item There exists a unique solution to the corresponding Lorentz equation, which is defined at any $\tau>\tau_0$ and tends to infinity not faster than $Me^{a\tau}$ at $\tau\rightarrow\infty$. Here, $\tau_0$, $M$ and $a>0$ are some constants;
  \item The perturbative series in $\la$ converges absolutely in a vicinity of the point $\la=0$ at sufficiently small $\la$;
\end{enumerate}
then a unique solution to Eq. \eqref{lde_integral} exists at sufficiently small $\la=\la_0>0$. If the value of $\la_0$ is smaller than the physical value of $\la$ then the physical solution to the LD equation at the physical value of $\la$ is obtained by an analytical continuation in $\la$.

A concrete prescription how to construct this analytical continuation depends on analytical properties of the solution to the Lorentz equation. For example, if this solution satisfies the first condition above and has a finite number of singularities in the part of the complex $\tau$-plane where $\re\tau>\tau_0$ and $\im\tau>0$, or $\re\tau>\tau_0$ and $\im\tau<0$ for some $\tau_0$, then $x^\mu(\la,\tau)$ has the same properties at sufficiently small $\la$. In that case, we can rotate the ray along which the integration contour tends to infinity so as to make the integral \eqref{lde_integral} convergent for any $a$. In particular, this procedure makes the right hand side of Eq. \eqref{lde_integral} convergent when we perturbatively solve Eq. \eqref{lde_integral} by Picard iterations starting with the solution to the Lorentz equation $x^\mu(0,\tau)$ satisfying the first condition above, while $3a\la\geq1$. Of course, the latter situation is rather unphysical since
\begin{equation}
    a\la\sim\al\frac{H}{H_0},\qquad H_0=\frac{m^2}{|e|\hbar},
\end{equation}
where $H$ is a characteristic value of the field strength, $\al$ is the fine structure constant and $H_0$ is the Schwinger field. However, we can define the physical solution even in this case.

Another, possibly more convenient, form of Eq. \eqref{lde_integral} can be derived if we write \cite{Plass}
\begin{equation}
    \dot{\ups}^2=\int_0^\infty dte^{-t}(\dot{\ups}f)(\tau+\la t/2),
\end{equation}
for physical solutions. Then
\begin{equation}\label{lde_integral1}
    \dot{\ups}_\mu(\tau)=\pr_\mu^\nu(\tau)\int_0^\infty dte^{-t}\left[f_\nu(\tau+\la t)+\la\int_0^tds(\dot{\ups}f)(\tau+\la s/2)\ups_\nu\left(\tau+\la(t-s)\right)\right].
\end{equation}
In this form, it is obvious that, in the absence of external fields, physical solutions to the LD equation are straight lines in the spacetime.

The integro-differential equations for physical solutions to the LD equation, which we have presented in this section, are not very useful in finding analytical solutions. But they are pertinent to numerical simulations, for example, by the Picard iterations, rather than the LD equation itself. Even if we set the initial conditions to their physical values and would solve numerically the LD equation then we shall obtain an unphysical runaway solution owing to machine approximation errors. Also, Eqs. \eqref{lde_integral} and \eqref{lde_integral1} are a good starting point for an investigation of the stochastic LD equation (see, e.g., \cite{JoHu,sd}). As in the case of numerical simulations, the unphysical solutions should be explicitly excluded to get rid of stochastically induced runaways. The study of this problem will be given elsewhere.

\section{Physical solutions for a planar motion}\label{linplan}

In this section, we study the planar motion of a charged particle obeying the LD equation. We call the motion of a particle planar (linear) provided that the trajectory of this particle in the space can be made planar (linear) by an appropriate Lorentz transform. For the linear motion, the worldline of a particle lies in a two-dimensional plane of the spacetime. For the planar motion, it lies in a three-dimensional hyperplane.  Throughout of this section, we mostly assume that the particle moves in the constant homogeneous external electromagnetic field, but some results can be generalized to a non-constant electromagnetic field of the special configuration. It will be mentioned in its place.

\subsection{Linear motion}\label{linmot}

Let us consider, at first, the hyperbolic motion, which is the motion with vanishing the LD force,
\begin{multline}\label{hyperbol_mot}
    \ddot{\ups}_\mu+\dot{\ups}^2\ups_\mu=0\;\Rightarrow\;\dot{\ups}^2=-\omega^2=const\;\Rightarrow\\
    \ups_\mu(\tau)=\al_\mu\cosh(\omega\tau)+\be_\mu\sinh(\omega\tau),\quad\al^2=-\be^2=1,\quad\al_\mu\be^\mu=0.
\end{multline}
The hyperbolic motion is the solution to the LD equation \eqref{lde} with a constant and homogeneous external electromagnetic field if, and only if,
\begin{equation}
    \ddot{\ups}^\mu=F^\mu_{\ \rho}F^\rho_{\ \nu}\ups^\nu=-\omega^2\ups^\mu.
\end{equation}
By the use of canonical forms (see \cite{DubNovFom} and also below) of the tensor $F_{\mu\nu}$, it is not difficult to show that the last equality (the equation on eigenvectors) is fulfilled if, and only if, there exists a Lorentz frame in which the particle moves along $3$-vectors of the electric and magnetic field strengths. In this system of coordinates, the problem reduces to a description of the linear motion.

For a linear motion, the LD equation \eqref{lde} is equivalent to (see, e.g., \cite{DeRaStTr,Plass,Dir} )
\begin{equation}\label{solution1d}
    \frac{\dot{\ups}}{\sqrt{1+\ups^2}}=\la\frac{d}{d\tau}\frac{\dot{\ups}}{\sqrt{1+\ups^2}}+E\;\Rightarrow\;\ups(\tau)=\sinh(c_2+E\tau+c_1e^{\la^{-1}\tau}).
\end{equation}
where $\ups:=\dot{x}(\tau)$. A generalization of this solution to the case of the electric field depending on $\tau$ is trivial. The solution \eqref{solution1d} with $E=0$ is also a general solution to the free LD equation written in the Lorentz frame, where the initial $3$-velocity and $3$-acceleration are parallel. The solution \eqref{solution1d} becomes physical if we take $c_1=0$. It is easy to verify that the solution \eqref{solution1d} satisfies the integro-differential equation \eqref{lde_integral} only at vanishing $c_1$. In order to make the integral convergent at $\la E\geq1$, we have to rotate the integration contour in Eq. \eqref{lde_integral} as it was described in the previous section.

\subsection{Planar motion}

\subsubsection{General consideration}\label{planmotgen}

Let us turn to the planar motion. When one of the Poincare-invariants of the electromagnetic field is not zero, the strength tensor \eqref{fmunu} can be represented as
\begin{equation}\label{strength_nondeg}
    F^{\mu\nu}=\omega_1e_0^{[\mu} e_1^{\nu]}+\omega_2e_2^{[\mu} e_3^{\nu]},\qquad (e_\al e_\be)=\eta_{\al\be},
\end{equation}
where $e_\al^\mu$, $\al=\overline{0,3}$, is a tetrad of eigenvectors of the tensor $(F^2)^\mu_\nu$. The eigenvalue $\omega^2_1$ of the tensor $(F^2)^\mu_\nu$ corresponds to the vectors $e^\mu_{0,1}$, and the eigenvalue $-\omega^2_2$ corresponds to the vectors $e^\mu_{2,3}$. In terms of the Poincare-invariants of the electromagnetic field $I_1=\mathbf{E}^2-\mathbf{H}^2$ and $I_2=2(\mathbf{EH})$, these omegas read as
\begin{equation}
    \omega^2_1=(\sqrt{I_1^2+I_2^2}+I_1)/2,\qquad\omega^2_2=(\sqrt{I_1^2+I_2^2}-I_1)/2.
\end{equation}
The Lorentz transforms, which do not change the strength tensor \eqref{strength_nondeg}, constitute the group $SO(1,1)\times SO(2)$. Its matrix representation is obvious.

In the degenerate case, when $I_1=I_2=0$, the strength tensor is given by
\begin{equation}\label{strength_deg}
    F^{\mu\nu}=\omega e_-^{[\mu}e_1^{\nu]},\qquad (e_ae_b)=\begin{bmatrix}
                                                      0 & 0 \\
                                                      0 & -1 \\
                                                    \end{bmatrix},
\end{equation}
where $e_a^\mu$, $a=\{-,1\}$, are eigenvectors of the tensor $(F^2)^\mu_\nu$ corresponding to zero eigenvalue. The normalized eigenvector $e_1^\mu$ is orthogonal to the vector $e_3^\mu$, which is a normalized eigenvector of $F^\mu_{\ \nu}$ and $(F^2)^\mu_\nu$ corresponding to zero eigenvalue. These conditions determine the vectors $e_1$ and $e_3$ uniquely up to adding the isotropic vector $e_-$ and inversion. The factor $\omega$ can be included to $e_-$, but we leave it in the expression so as to control the external field. The strength tensor \eqref{strength_deg} is invariant with respect to the two-dimensional Abelian subgroup of the Lorentz group generated by the two elements
\begin{equation}
    \La_1^{\mu\nu}=\eta^{\mu\nu}+r_1e_3^{[\mu}e_-^{\nu]}+\frac{r_1^2}2e_-^\mu e_-^\nu,\qquad \La_2^{\mu\nu}=\eta^{\mu\nu}+r_2e_1^{[\mu}e_-^{\nu]}+\frac{r_2^2}2e_-^\mu e_-^\nu,
\end{equation}
where $r_1$ and $r_2$ are the group parameters. Thus, in both degenerate and non-degenerate cases, we anticipate two integrals of motion of the LD equation provided the eigenvectors of the tensors $F^\mu_{\ \nu}$ and $(F^2)^\mu_\nu$ do not depend on a point of the spacetime.

In order to obtain these integrals of motion, it is useful to introduce new variables adjusted to the action of the symmetry group. In the non-degenerate case, they are
\begin{equation}\label{subs_eh}
    \mathrm{v}_0=:\sqrt{u_e}\cosh\psi,\qquad \mathrm{v}_1=:\sqrt{u_e}\sinh\psi,\qquad \mathrm{v}_2=:\sqrt{u_h}\sin\vf,\qquad \mathrm{v}_3=:\sqrt{u_h}\cos\vf,
\end{equation}
where $\mathrm{v}_\al:=(e_\al\ups)$, $u_e-u_h=1$, and $\vf$ and $\psi$ are the symmetry group parameters. Then, convolving the LD equation \eqref{lde} with the eigenvectors, we have
\begin{equation}\label{integrals_mot}
\begin{split}
    u_e(\tau)\dot{\psi}(\tau)&=c_1e^{\la^{-1}\tau}+\int_\tau^\infty\frac{dt}{\la}e^{-\la^{-1}(t-\tau)}\omega_1 u_e(t),\\ u_h(\tau)\dot{\vf}(\tau)&=c_2e^{\la^{-1}\tau}-\int_\tau^\infty\frac{dt}{\la}e^{-\la^{-1}(t-\tau)}\omega_2 u_h(t).
\end{split}
\end{equation}
In the degenerate case, the analogous variables read as
\begin{equation}\label{subs_iso}
    \mathrm{v}_1=:\mathrm{v}_-r_2,\qquad \mathrm{v}_3=:\mathrm{v}_-r_1,\qquad \mathrm{v}_+=:\mathrm{v}_-^{-1}+\mathrm{v}_-(r_1^2+r_2^2),
\end{equation}
where $e_+^\mu$ is an isotropic vector orthogonal to $e_1^\mu$ and $e_3^\mu$, and such that $(e_+e_-)=2$. The respective integrals of motion become
\begin{equation}\label{integrals_mot1}
\begin{split}
    u(\tau)\dot{r}_1(\tau)&=c_1e^{\la^{-1}\tau},\\
    u(\tau)\dot{r}_2(\tau)&=c_2e^{\la^{-1}\tau}+\int_\tau^\infty\frac{dt}{\la}e^{-\la^{-1}(t-\tau)}\omega u(t),
\end{split}
\end{equation}
where $u(\tau):=\mathrm{v}_-^2(\tau)$. The non-vanishing constants $c_1$ and $c_2$ in Eqs. \eqref{integrals_mot} and \eqref{integrals_mot1} correspond to unphysical solutions. So, they should be set to zero and we do not take them into account henceforth.

Consider the particular case $I_2=0$ (see \cite{GuptaEl}). If $I_1>0$ then $\omega_2$ is zero and we get
\begin{equation}
    \mathrm{v}_2=\mathrm{v}_3=0\quad\text{or}\quad\dot{\vf}=0.
\end{equation}
The first case is a linear motion, the second case is a planar one. When $I_1<0$, we have $\omega_1=0$ and
\begin{equation}
    \dot{\psi}=0.
\end{equation}
In the degenerate case $I_1=I_2=0$,
\begin{equation}
    \dot{r}_1=0.
\end{equation}
By definition of the variables $\psi$ and $r_1$ as the group parameters, the particle can be confined to a plane by an appropriate Lorentz transform in these cases.

Thus we have proved the following statement. If $I_2=0$ and the field strength tensor admits the representation \eqref{strength_nondeg} or \eqref{strength_deg} with the constant eigenvectors $e_\al^\mu$ then a charged particle obeying the LD equation executes a planar motion. In a constant homogeneous electromagnetic field, the converse statement is also true. Namely, if a charged particle obeying the LD equation executes an essentially planar (i.e. non linear) motion then $I_2=0$.

\subsubsection{Second order equation}\label{planmotsecor}

Now we investigate the planar motion in detail. Let us characterize this motion by a tetrad
\begin{equation}
    e^\mu_\al \eta_{\mu\nu}e^\nu_\be=\eta_{\al\be},\qquad e^\mu_3 e^\al_\mu=0,\qquad (e_3)^2=-1,
\end{equation}
where the indices $\al$ and $\be$ run the values $0,1,2$. They are risen and lowered by the metric $\eta_{\al\be}=diag(1,-1,-1)$. The worldline of the particle and the external electromagnetic field admit a representation
\begin{equation}
\begin{gathered}
    \ups^\mu(\tau)=\ups^\al(\tau)e_\al^\mu,\qquad e^\mu_3\ups_\mu(\tau)=0,\\
    F_{\mu\nu}=f_{\al\be}e^\al_\mu e^\be_\nu,\qquad f_{\al\be}=\omega\e_{\al\be\ga}\xi^\ga,
\end{gathered}
\end{equation}
where $\e_{012}=1$ and $\xi^2=\{\pm1,0\}$. The LD equation \eqref{lde} is rewritten as (for the Lorentz equation see, e.g., \cite{Plyushch})
\begin{equation}\label{lde tetr}
    \dot{\ups}_\al=\omega\e_{\al\be\ga}\ups^\be \xi^\ga+\la(\ddot{\ups}_\al+\dot{\ups}^2\ups_\al),\qquad\ups^2=1.
\end{equation}
We see that the LD equation of a charged particle confined to a plane possesses a symmetry. This is a residue of the symmetry discussed above after the reduction to the plane. The residue symmetry group is constituted by the Lorentz transforms leaving the vector $\xi^\al$ intact. So, if $\xi^2\leq0$ this symmetry group is isomorphic to $SO(1,1)$, and if $\xi^2>0$ it is isomorphic to $SO(2)$. This symmetry allows us to reduce the problem of integration of the system of equations \eqref{lde tetr} to an integration of the autonomous system of three first order equations or the one second order equation.

To this end, we introduce new more convenient variables
\begin{equation}
    m^\al:=\e^{\al\be\ga}\dot{\ups}_\be\ups_\ga.
\end{equation}
In these variables, the LD equation \eqref{lde tetr} turns into a system of the first order equations
\begin{equation}\label{lde in m}
    \la\dot{m}_\al=m_\al+\omega(\xi_\al-p\ups_\al),\qquad\dot{\ups}_\al=-\e_{\al\be\ga}m^\be\ups^\ga,\qquad m_\al\ups^\al=0,\qquad\ups^2=1,
\end{equation}
where $p:=\xi_\al\ups^\al$. Then we introduce the invariants of the symmetry group action
\begin{equation}\label{invar}
\begin{aligned}
    a&=\xi_\al m^\al,&\qquad b&=p^{-1}\e^{\al\be\ga}\xi_\al m_\be\ups_\ga=-p^{-1}\dot{p},\\
    s&=p^{-2}(m^2-\xi^2a^2),&\qquad u&=p^2-\xi^2.
\end{aligned}
\end{equation}
These invariants are dependent. From their definition, it is not difficult to obtain the identity
\begin{equation}\label{ident}
    a^2\frac{1+\xi^2u}{\xi^2+u}+b^2=-su.
\end{equation}
The invariants \eqref{invar} evolve according to the equations
\begin{equation}\label{invar_evol0}
\begin{aligned}
    \la\dot{a}&=a-\omega u,&\qquad\dot{u}&=-2b(\xi^2+u),\\
    \la\dot{s}&=2s(1+\la b)+2\omega a\frac{1+\xi^2u}{\xi^2+u},&\qquad
    \la\dot{b}&=b+\la\xi^2s+\la a^2\frac{|\xi^2|-1}{u}=b-\la\frac{a^2+\xi^2b^2}{u}.
\end{aligned}
\end{equation}
The first equation in this system is the one of the equations of the system \eqref{integrals_mot} or \eqref{integrals_mot1} written in a differential form. The physical solutions are described by the one integro-differential equation on $u(\tau)$. It is obtained from the above equations if we write
\begin{equation}
    a(\tau)=\int_\tau^\infty\frac{dt}{\la}e^{-\la^{-1}(t-\tau)}\omega u(t),\qquad p^2(\tau)s(\tau)=\int_\tau^\infty\frac{2dt}{\la}e^{-2\la^{-1}(t-\tau)}\omega p^2(t)a(t),
\end{equation}
and substitute these expressions to the identity \eqref{ident} with the function $b(\tau)$ taken from the second equation of the system \eqref{invar_evol0}. The solution of this integro-differential equation is specified by the only one arbitrary constant. If this solution or some unphysical solution to the system \eqref{invar_evol0} are known, we can integrate the LD equation.

Indeed, in a general position the vector $\dot{\ups}_\al$ can be expressed as a linear combination of the vectors $\xi_\al$, $\ups_\al$ and $\e_{\al\be\ga}\xi^\be\ups^\ga$. The coefficients of this decomposition are certain functions of the invariants which are already known. Therefore, we need to integrate a system of linear equations with variable coefficients. So,
\begin{equation}\label{accelera}
    \dot{\ups}_\al=\frac{a}{u}\e_{\al\be\ga}\xi^\be\ups^\ga+\frac{bp}{u}(\xi_\al-p\ups_\al).
\end{equation}
Because of the orthogonality condition, only two equations are independent. Now we make a substitution to Eq. \eqref{accelera} of the form \eqref{subs_eh}, \eqref{subs_iso}. In the case $\xi^\al=(1,0,0)$, we have
\begin{equation}\label{subs h}
    \ups_0=p,\qquad\ups_1=\sqrt{u}\cos\vf,\qquad\ups_2=\sqrt{u}\sin\vf,
\end{equation}
and
\begin{equation}\label{freq_H}
    u\dot{\vf}=-a,
\end{equation}
whence the momenta $\ups_\al$ are found in quadratures. If $\xi^\al=(0,0,1)$ then we substitute
\begin{equation}\label{subs e}
    \ups_0=\sqrt{u}\cosh\psi,\qquad\ups_1=\sqrt{u}\sinh\psi,\qquad\ups_2=p.
\end{equation}
This results in
\begin{equation}\label{freq_E}
    u\dot{\psi}=a.
\end{equation}
In the third case $\xi^\al=(1,0,1)$, we do
\begin{equation}\label{isotropic}
    \ups_1=rp,\qquad\ups_-=p,\qquad p\ups_+=1+ur^2,
\end{equation}
and arrive at
\begin{equation}\label{freq_HE}
    u\dot{r}=a.
\end{equation}
The case of an arbitrary vector $\xi^\al$ is reduced to the considered ones by a proper Lorentz transform of the tetrad indices. Notice that the equations of motion \eqref{invar_evol0}, \eqref{freq_H}, \eqref{freq_E} and \eqref{freq_HE} are valid for a non-constant external field parameter $\omega$. In accordance with our general considerations, physical solutions are specified by four constants -- two constants specify the initial position on the plane, one determines $u(\tau)$ and another one is needed to pick out the unique solution from Eqs. \eqref{freq_H}, \eqref{freq_E} or \eqref{freq_HE}.

Thus, we have to find the evolution of invariants described by Eqs. \eqref{invar_evol0}. In case of a constant $\omega$, the autonomous system \eqref{invar_evol0} is equivalent to the one second order differential equation on the function $a(u)$ or its inverse $u(a)$
\begin{equation}
\begin{split}
    a''&=-\frac{[2au+\xi^2(a+\omega u)]a'}{2u(a-\omega u)(\xi^2+u)}-\frac{2\la^2 a^2(\xi^2+u)a'^3}{u(a-\omega u)^2},\\
    \ddot{u}&=\frac{[2au+\xi^2(a+\omega u)]\dot{u}^2}{2u(a-\omega u)(\xi^2+u)}+\frac{2\la^2 a^2(\xi^2+u)}{u(a-\omega u)^2}.
\end{split}
\end{equation}
Even in the simplest case $\xi^2=0$, when these equations can be cast into the form
\begin{equation}
    a''=-\frac{aa'}{u(a-u)}-\frac{2a^2a'^3}{(a-u)^2},\qquad\ddot{u}=\frac{a\dot{u}^2}{u(a-u)}+\frac{2a^2}{(a-u)^2},
\end{equation}
we have not succeeded in finding a general solution.

\subsubsection{Asymptotics}\label{planmotasym}

However, we can investigate the asymptotics of exact physical solutions to the LD equation at large times. It is easily done in the coordinates where the vector field of the system \eqref{invar_evol0} has no singularities. Making a change of variables
\begin{equation}
    a=u\bar{a},\qquad b=u\bar{b},
\end{equation}
we come to
\begin{equation}\label{invar evol 1}
    \lambda\dot{\bar{a}}=\bar{a}[1+2\la\bar{b}(\xi^2+u)]-\omega,\qquad\lambda\dot{\bar{b}}=\bar{b}-\la[\bar{a}^2-\bar{b}^2(\xi^2+2u)],\qquad\dot{u}=-2\bar{b}u(\xi^2+u).
\end{equation}
It is not difficult to find the stationary points of this system. We are interested only in physical solutions and physical stationary points. A physical stationary point as a particular case of a physical solution should be regular in $\la$. Again we have three cases.

The case $\xi^2=1$ is a planar motion in a constant homogeneous magnetic field \cite{Plass,Endres}. The system \eqref{invar evol 1} has two stationary points, one of them being physical
\begin{equation}\label{stationary_h}
    \bar{a}=\frac{\omega}{g},\qquad\bar{b}=\frac{g-1}{2\la},\qquad u=0,\qquad g:=2^{-1/2}\left(1+\sqrt{1+16\la^2\omega^2}\right)^{1/2}.
\end{equation}
Linearizing the system \eqref{invar evol 1} in a vicinity of this point, we obtain the asymptotics of the exact solution to the LD equation
\begin{equation}\label{solution_H}
\begin{split}
    \de\bar{a}&=-u(0)Ae^{\la^{-1}(1-g)\tau}+e^{\la^{-1}g\tau}\left[(c_1+u(0)A)\cos\frac{2\omega\tau}{g}+(c_2+u(0)B)\sin\frac{2\omega\tau}{g}\right],\\
    \de\bar{b}&=-u(0)Be^{\la^{-1}(1-g)\tau}+e^{\la^{-1}g\tau}\left[(c_2+u(0)B)\cos\frac{2\omega\tau}{g}-(c_1+u(0)A)\sin\frac{2\omega\tau}{g}\right],\\
    \de u&=u(0)e^{\la^{-1}(1-g)\tau},\qquad A:=\frac{\omega(g-1)}{g(5g-4)},\qquad B:=\frac{3(g-1)^2}{2\la(5g-4)}.
\end{split}
\end{equation}
The terms in the brackets describe runaway solutions. They are unphysical and have to be set to zero by a proper choice of the initial conditions. Then the physical solution to Eqs. \eqref{invar evol 1} is solely specified by the initial condition on $u$. The first term in the first line in Eqs. \eqref{solution_H} describes a correction to the rotational speed of a charged particle due to the radiation reaction. Inasmuch as $u$ is non-negative, this correction has an opposite sign with respect to the main contribution, i.e., the rotational speed increases with time and tends exponentially to its limiting value \eqref{stationary_h}. The limiting value is, of course, lesser than the cyclotron frequency. In case at hand, $u$ is related to the kinetic energy of the particle and so the expression for $\de u $ in \eqref{solution_H} describes its decreasing.

The case $\xi^2=-1$ corresponds to a planar motion in a constant homogeneous electric field. The system \eqref{invar evol 1} possesses two stationary points. Only one of these points is physical
\begin{equation}\label{stationary_e}
    \bar{a}=\omega,\qquad\bar{b}=\frac{g-1}{2\la},\qquad u=1,\qquad g:=\sqrt{1+4\la^2\omega^2}.
\end{equation}
The linearized in a neighbourhood of this point LD equation has the solution
\begin{equation}\label{solution_E}
\begin{split}
    \de\bar{a}&=-\omega u(0)(1-g^{-1})e^{\la^{-1}(1-g)\tau}+e^{\la^{-1}\tau}\left[c_1+\omega u(0)(1-g^{-1})\right],\\
    \de\bar{b}&=-u(0)Be^{\la^{-1}(1-g)\tau}+\frac{2\la \omega}{g-1}e^{\la^{-1}\tau}\left[c_1+\omega u(0)(1-g^{-1})\right]+e^{\la^{-1}g\tau}\left[c_2-\frac{2c_1\la\omega}{g-1}+\frac{u(0)(g-1)}{\la(1-2g)}\right],\\
    \de u&=u(0)e^{\la^{-1}(1-g)\tau},\qquad B:=\frac{(g-1)^2(2g+1)}{2\la g(2g-1)}.
\end{split}
\end{equation}
Unphysical solutions are the terms in the square brackets. Demanding their vanishing, we uniquely determine the integration constants $c_1$ and $c_2$ through $u(0)$. The correction to the ``frequency'' $\dot{\psi}$ increases with time and tends exponentially to the limiting value \eqref{stationary_e}. The expression for $\de u(0)$ in \eqref{solution_E} describes an evolution of the square of the momentum component normal to the electric field. As expected, this component exponentially tends to zero and the solution passes into the hyperbolic motion \eqref{hyperbol_mot}.

The case $\xi^2=0$ is more involved. To shorten formulas, we redefine the variables entering \eqref{invar evol 1}
\begin{equation}\label{replacement}
    \bar{a}\rightarrow \omega\bar{a},\qquad\bar{b}\rightarrow\la\omega^{2}\bar{b},\qquad u\rightarrow(\la\omega)^{-2}u,\qquad\tau\rightarrow\la\tau,
\end{equation}
and shall restore the original notation where it will be necessary to make estimations. After this redefinition, a regularity in $\la$, which distinguishes physical solutions, means a regularity of the solution in $\tau^{-1}$. Then the system \eqref{invar evol 1} has a single stationary point
\begin{equation}\label{crit_point}
    \bar{a}=1,\qquad\bar{b}=1,\qquad u=0.
\end{equation}
This point is degenerate and, therefore, the solutions to the linearized system improperly describe a behaviour of solutions to the LD equation in a vicinity of this point. To obtain a correct asymptotics, we integrate the last equation in \eqref{invar evol 1}
\begin{equation}
    u=u(0)\left[1+2u(0)\tau+2u(0)\int_0^\tau dt\de\bar{b}(t)\right]^{-1}.
\end{equation}
The integrand of the third term in the square brackets tends to zero. Consequently, the second term in the square brackets will dominate at large times
\begin{equation}\label{assumptions}
    \tau\gg\la,\qquad\tau\gg(2\la\omega^2u(0))^{-1},
\end{equation}
and we can take
\begin{equation}
    u\approx\tau^{-1}/2.
\end{equation}
Then the equations for the leading asymptotics read as
\begin{equation}\label{linearized_he}
    \de\dot{\bar{a}}=\de\bar{a}+\tau^{-1}\de\bar{b}+\tau^{-1},\qquad\de\dot{\bar{b}}=-2\de\bar{a}+\de\bar{b}+\tau^{-1},
\end{equation}
where we keep only the leading terms. The system \eqref{linearized_he} possesses runaway solutions, which are nonregular in $\la$ ($\tau^{-1}$) and, consequently, unphysical. They can be removed by an appropriate choice of the initial data. As for the physical solutions, their leading asymptotics takes the form
\begin{equation}\label{asympt}
\begin{split}
    \bar{a}&=1-\tau^{-1}-\tau^{-2}(3\ln\tau+2u(1)-6)+o(\tau^{-2})=1-\frac1{\tau}\left(\frac{e^{-5/3}\tau}{8u^3(1)}\right)^{3/\tau}+o(\tau^{-2}),\\
    \bar{b}&=1-3\tau^{-1}-\tau^{-2}(9\ln\tau+6u(1)-23)+o(\tau^{-2})=1-\frac3{\tau}\left(\frac{e^{-20/9}\tau}{8u^3(1)}\right)^{3/\tau}+o(\tau^{-2}),\\
    u&=\frac12\left[\tau^{-1}+\tau^{-2}(3\ln\tau+2u(1)-1)\right]+o(\tau^{-2})=\frac1{2\tau}\left(\frac{\tau}{8u^3(1)}\right)^{3/\tau}+o(\tau^{-2}).
\end{split}
\end{equation}
Here we also add a next to leading correction to the asymptotics, which can be derived from the initial non-linear system \eqref{invar evol 1}. The last equalities in \eqref{asympt} show how the power of the proper time entering the asymptotics tends to its limiting value. The initial value $u(1)$ in these last formulas differs from $u(1)$ appearing in the second equalities in \eqref{asympt}. These initial values are related in an evident manner.

Substituting the asymptotics \eqref{asympt} into Eqs. \eqref{isotropic}, \eqref{freq_HE} and bearing in mind the replacement \eqref{replacement}, we find
\begin{equation}
\begin{split}
    \ups_1&=\left(\frac{\tau}{2\la}\right)^{1/2}\left(2\ups^2_1(\la)\frac{\tau}{\la}\right)^{\la/2\tau}+o\left((\la/\tau)^{1/2}\right),\\
    \ups_0&=\la\omega\left(\frac{\tau}{2\la}\right)^{3/2}\left(\frac{\la\omega^2\tau}{8\ups_0^2(\la)}\right)^{-\la/2\tau}+o\left((\tau/\la)^{1/2}\right)=-\ups_2.
\end{split}
\end{equation}
As we see, the system goes to the universal regime. For example, the quantity
\begin{equation}\label{ratio}
    \frac{\ups_0}{\ups_1^3}\approx-\frac{\ups_2}{\ups_1^3}\approx\la\omega\left(\frac{\omega\ups_1^3(\la)\tau^2}{\la\ups_0(\la)}\right)^{-\la/\tau}
\end{equation}
ceases to depend on the initial data and tends to $\la\omega$. This occurs on the proper time scales
\begin{equation}
    \tau\gg\la\left|\ln\frac{\ups_0(\la)}{\la\omega\ups_1^3(\la)}\right|.
\end{equation}
This effect is essentially due to the radiation reaction. After the lapse of a certain time, the charged particles moving on the plane in the electromagnetic field with the invariants $I_1=I_2=0$ will have the same ratio of momenta of the form \eqref{ratio} irrespective of their initial momenta. So, if we measure the momenta components of these charged particles, the measured data will lie on the cubic parabola determined by Eq. \eqref{ratio}.

For comparison, we give here the well-known solution to the Lorentz equation in this electromagnetic field
\begin{equation}
\begin{gathered}
    \ups_1=\ups_1(0)+\sqrt{u(0)}\omega\tau,\qquad\ups_0=\ups_0(0)+\ups_1(0)\omega\tau+\sqrt{u(0)}\frac{\omega^2\tau^2}2,\\
    \ups_2=\ups_2(0)-\ups_1(0)\omega\tau-\sqrt{u(0)}\frac{\omega^2\tau^2}2.
\end{gathered}
\end{equation}
In this case, the quantity analogous to \eqref{ratio}, which tends to a constant at large times, is the ratio
\begin{equation}
    \frac{\ups_0}{\ups_1^2}\approx-\frac{\ups_2}{\ups_1^2}\approx\frac1{2\sqrt{u(0)}}.
\end{equation}
Its limiting value depends on the initial conditions. The ratio \eqref{ratio} goes to zero as $\tau^{-1}$ with asymptotics depending on the initial data.

\subsubsection{Landau-Lifshitz equation}\label{planmotLL}

Now we investigate the planar motion of a charged particle in a constant homogeneous electromagnetic field in the framework of the so-called Landau-Lifshitz equation \cite{LandLifsh}. This is an approximate equation describing the physical solutions to the LD equation. It is obtained from the LD equation by the reduction of order procedure, while $\la$ being assumed to be a small parameter. Thus, let us seek for solutions to the LD equation \eqref{lde in m} in a class of functions which
\begin{equation}
    \left|\frac{d^km_\al}{d\tau^k}\right|=O(1),\qquad\left|\frac{d^k\ups_\al}{d\tau^k}\right|=O(1),\qquad k=\overline{0,\infty},
\end{equation}
with respect to the small parameter $\la$. Also, we restrict ourself by the first correction in $\la$ to the Lorentz equation.

Differentiating the first equation in \eqref{lde in m} with respect to $\tau$, we find $\dot{m}_\al$. Then we substitute it to the initial equation and come to
\begin{equation}
    m_\al=-\omega(\xi_\al-p\ups_\al-\la\dot{p}\ups_\al-\la p\dot{\ups}_\al).
\end{equation}
By the use of this relation, the second equation in \eqref{lde in m} describing the evolution of the momentum $\ups_\al$ has the form
\begin{equation}
    \dot{\ups}_\al=\omega\e_{\al\be\ga}\xi^\be\ups^\ga+\la\omega^2p(\xi_\al-p\ups_\al),
\end{equation}
where we neglect the higher orders in $\la$. Convolving this equation with $\xi_\al$, we arrive at
\begin{equation}
    \dot{u}=-2\la \omega^2u(\xi^2+u),
\end{equation}
whence
\begin{equation}
    u=\frac{\xi^2u(0)}{(\xi^2+u(0))e^{2\la\omega^2\xi^2\tau}-u(0)},\quad\xi^2=\pm1;\qquad u=\frac{u(0)}{1+2\la\omega^2u(0)\tau},\quad\xi^2=0.
\end{equation}
If $\xi^\al=(1,0,0)$, the substitution \eqref{subs h} gives
\begin{equation}
    \dot{\vf}=-\omega.
\end{equation}
If $\xi^\al=(0,0,1)$, the substitution \eqref{subs e} results in
\begin{equation}
    \dot{\psi}=\omega.
\end{equation}
These formulas are in agreement with Eqs.  \eqref{solution_H} and \eqref{solution_E} up to the first order in $\la$. In the case $\xi^\al=(1,0,1)$, we have
\begin{equation}
\begin{gathered}
    \ups_1=\frac{\ups_1(0)+\sqrt{u(0)}\omega\tau}{\sqrt{1+2\la\omega^2u(0)\tau}},\qquad\ups_0+\ups_2=\frac{\sqrt{u(0)}}{\sqrt{1+2\la \omega^2u(0)\tau}},\\
    \ups_0-\ups_2=\frac{\ups_0(0)-\ups_2(0)+2\omega(\ups_1(0)+\la\omega\sqrt{u(0)})\tau+\sqrt{u(0)}\omega^2\tau^2}{\sqrt{1+2\la\omega^2u(0)\tau}}.
\end{gathered}
\end{equation}
We see that the system passes to the universal regime at sufficiently large proper times
\begin{equation}
    \ups_1\approx\frac{\tau^{1/2}}{\sqrt{2\la}},\qquad\ups_0+\ups_2\approx \omega^{-1}\frac{\tau^{-1/2}}{\sqrt{2\la}},\qquad\ups_0-\ups_2\approx \omega\frac{\tau^{3/2}}{\sqrt{2\la}}.
\end{equation}
The limiting value of the ratio \eqref{ratio} is the same for these solutions as for the physical solutions to the exact LD equation. Thus, the Landau-Lifshitz equation correctly reproduces the asymptotics of the physical solution at large $\tau$ in spite of the fact that this asymptotics is not regular in $\la$.

\section{Discussion}\label{discuss}

In this paper, we have investigated a planar motion of charged particles obeying the LD equation. We gave a detailed study of asymptotics of the planar physical solutions to the LD equation in a constant homogeneous electromagnetic field. One of the main results of our study is an interesting asymptotics of the physical solution to the LD equation in this electromagnetic field with vanishing invariants $I_1=I_2=0$. According to the classical radiation reaction theory, the charged particles moving on the plane in such the electromagnetic field must have the same ratio \eqref{ratio} of the momenta components at sufficiently large times. The existence of this asymptotics can be verified in a purely quantum electrodynamical context.

Namely, the Dirac equation in that electromagnetic field can be exactly solved \cite{BagGit}. Then, by the use of this complete set of solutions, we construct quantum electrodynamics on the given background \cite{GFSh} and define the $S$-matrix. The operators of covariant momenta $\mathcal{P}_0=m\ups_0$ and $\mathcal{P}_1=m\ups_1$ commute. Therefore, we can construct a complete set of their eigenfunctions and evaluate the transition amplitudes to these states. For the one-photon radiation amplitude, the transition probability reads as
\begin{equation}\label{trans_prob}
    \sum_{k,\la} \lan\be|\hat{U}^\dag|\mathcal{P}_0,\mathcal{P}_1;k,\la\ran \lan\mathcal{P}_0,\mathcal{P}_1;k,\la|\hat{U}|\be\ran,
\end{equation}
where $\hat{U}$ is the evolution operator over an infinite time, $\be$ are the quantum numbers characterizing the initial state of the electron, and $k$ and $\la$ are the momentum and polarization of the radiated photon. Provided the classical radiation reaction theory is viable, there must exist a range of the quantum numbers $\be$ and strengths of the electromagnetic field such that the transition probability \eqref{trans_prob} is mostly concentrated on the cubic parabola
\begin{equation}
    \mathcal{P}_0=\frac{2e^2 E}{3m^4}\mathcal{P}_1^3.
\end{equation}
Of course, using formula \eqref{trans_prob}, we disregard the multiple photon production and the production of the electron-position pairs, but these contributions are proportional to the fine structure constant and neglible for reasonable strengths of the electromagnetic field. We postpone a detail study of this effect in quantum electrodynamics for a future research.

Also note that, as it was pointed out in \cite{Bhabha}, the notion of a physical solution seems to be quite general and can be exploited to give a proper interpretation of the higher-order derivative corrections to effective actions in quantum field theory (see, e.g., \cite{JaLloMol,Woodart,Moroz} and also their quantization in \cite{LyakhovichPhD,Plyushch1}). Usually, these terms arise from the heat kernel expansion over the regularization parameter $\La$ or large mass $m$ of the one-loop correction \cite{VasilHeatKer} whereas the higher terms of this expansion are casted out. By analogy with the considerations presented in Appendix, we should demand a regularity of solutions to the effective equations of motion in the small expansion parameter -- the inverse regularization parameter or inverse large mass. Then a neglect of the nonlocal reminder of the effective action is justified. In many instances, this regularity can be related to the regularity with respect to the coupling constants. Though, it does not mean that we assume a smallness of the couplings or these higher-order derivative terms. As far as the heat kernel expansion is concerned, one can distinguish the corrections of three types: i) the divergent higher derivative terms like, for example, in $R^2$ gravity; ii) the higher derivative terms disappearing in the regularization removal limit; iii) the higher derivative terms resulting from the large mass expansion of a finite part of the effective action. In these cases, the coefficients at the higher derivative terms have a form
\begin{equation}
    i)\;\bar{\la}_a^{-1}+e_a(\bar{\la})/f_a(m/\La),\qquad ii)\;e_b(\bar{\la})f_b(m/\La),\qquad iii)\;e_c(\bar{\la})f_c(R/m^2),
\end{equation}
where $e$ and $f$ are some functions going to zero at the origin, $R$ schematically denotes the field strengths or their derivatives. The coefficients in the first case are the renormalized inverse couplings $\la_a^{-1}$. Now it is easy to see that a regularity of the expression in the coupling constants $\la_a$ implies its regularity in $\La^{-1}$. As for the large mass expansion, we additionally have to require a regularity of solutions to the effective equations of motion in $m^{-1}$. Just to demonstrate what we mean, consider the effective action with a higher derivative correction coming from the (self)interaction
\begin{equation}
    S[\phi]=\frac12\int dx\phi(-\Box-m^2+\la\Box^2)\phi,
\end{equation}
where $\la$ is proportional to the coupling constants or, possibly, to the inverse large mass. The physical sector of this model is equivalent to
\begin{equation}
    S_{phys}[\phi]=\frac12\int dx\phi\left[-\Box-(\sqrt{1+4\la m^2}-1)/2\la\right]\phi.
\end{equation}
As long as as the coupling constants are scalars with respect to a symmetry group of the model, its physical sector possesses this symmetry as well. An elimination of the unphysical sector in free models is a simple task, but, for full interacting models, such an explicit elimination becomes complicated and results in an appearance of the nonlocal projectors to the physical states in the effective action.

\appendix

\section{Regularity condition in Maxwell electrodynamics}

In this appendix, we show that the solutions to the coupled system of Maxwell-Lorentz equations are regular in the coupling constants before taking a regularization removal limit.

Consider a model with the action functional \eqref{action particl}. The effective equations of motion of a charged particle with a bare mass $\bar{m}$ look like
\begin{equation}\label{lde_int_gen}
    \bar{m}\ddot{x}_\mu(\tau)=eF_{\mu\nu}(x(\tau))\dot{x}^\nu(\tau)+4e^2\int d\tau'\theta(X^0(\tau,\tau'))\de'(X^2(\tau,\tau')-\e)X_{[\mu}(\tau,\tau')\dot{x}_{\nu]}(\tau')\dot{x}^\nu(\tau),
\end{equation}
where $X_\mu(\tau,\tau'):=x_\mu(\tau)-x_{\mu}(\tau')$, and we have used the regularization of the retarded Green function of the form
\begin{equation}\label{green_func_reg_0}
    G^{-}(x)=\frac{\theta(x^0)}{2\pi}\de(x^2)\;\rightarrow\;G^{-}_\e(x)=\frac{\theta(x^0)}{2\pi}\de(x^2-\e),
\end{equation}
where $\e$ is a regularization parameter. We shall assume that $F_{\mu\nu}(x)$ is smooth and bounded on the spacetime, $F_{\mu\nu}(x(\tau))$ vanishes at $\tau<\tau_0$ for some $\tau_0$, and the solution to Eq. \eqref{lde_int_gen} has the following asymptotics in the past
\begin{equation}\label{asympt_past}
    x_\mu(\tau)=(\tau-\tau_0)\ups_\mu+\bar{x}_\mu(\tau),\qquad\ups^2=1,
\end{equation}
where $\bar{x}_\mu(\tau)$ is zero at $\tau<\tau_0$, and $\ups_\mu$ is a constant $4$-vector. The unphysical solutions may appear on the scales of a classical electron radius. Therefore, we are looking for the solution of the form
\begin{equation}\label{subs_irregul}
    x_\mu(\tau)=\bar{r}_e y_\mu(\tau/\bar{r}_e),\qquad \bar{r}_e:=e^2/\bar{m}.
\end{equation}
The function $y_\mu(s)$ also depends on other dimensionless combinations of parameters entering Eq. \eqref{lde_int_gen}, but we do not write them explicitly. Upon substitution \eqref{subs_irregul}, we arrive at
\begin{equation}\label{lde_int_gen_y}
    \ddot{y}_\mu(s)=\frac{e}{\bar{m}}\bar{r}_eF_{\mu\nu}(y(s))\dot{y}^\nu(s)+4\int ds'\theta(Y^0(s,s'))\de'(Y^2(s,s')-\bar{\e})Y_{[\mu}(s,s')\dot{y}_{\nu]}(s')\dot{y}^\nu(s),
\end{equation}
where $\bar{\e}:=\e/\bar{r}_e^2$ and other evident redefinitions have been done. The integral in this equation can be taken
\begin{equation}\label{I}
    I=\left[\frac{Y_{[\mu}(s,s')\ddot{y}_{\nu]}(s,s')}{(\dot{y}_\rho(s')Y^\rho(s,s'))^2}+\frac{Y_{[\mu}(s,s')\dot{y}_{\nu]}(s')}{(\dot{y}_\rho(s')Y^\rho(s,s'))^3}(1-\ddot{y}_\rho(s')Y^\rho(s,s'))\right]\dot{y}^\nu(s),
\end{equation}
where $s'$ is determined by the conditions $Y^2(s,s')=\bar{\e}$, $Y^0(s,s')<0$. The LD equation is obtained by expanding Eqs. \eqref{lde_int_gen_y} and \eqref{I} in the asymptotic series in $\e$ around zero and discarding the terms vanishing at $\e\rightarrow0$. In view of the asymptotic behaviour of the solution \eqref{asympt_past}, the asymptotics of the integral $I$ at small $\bar{r}_e$ readily follows
\begin{equation}\label{I1}
    I\underset{\bar{r}_e\rightarrow0}{\rightarrow}\e^{-3/2}\bar{r}_e^3\bar{y}_{[\mu}(s)\ups_{\nu]}\dot{y}^\nu(s).
\end{equation}
Hence, in this limit, the integro-differential equation \eqref{lde_int_gen} reduces the Lorentz-type differential equation with the effective strength tensor: $F_{\mu\nu}$ plus the correction \eqref{I1}. According to the general theorems of ordinary differential equations theory, the solutions to this equation are regular in $\bar{r}_e$. The effective strength tensor tends to zero when $\bar{r}_e\rightarrow0$, and, in this limit, the particle moves along a straight line
\begin{equation}
    \ddot{y}_\mu(s)=0.
\end{equation}
Restituting the notation \eqref{subs_irregul}, we see that the limiting trajectory of the particle is indeed regular in $\bar{r}_e$. If we compel the Lorentz force to be constant at $\bar{r}_e\rightarrow0$ increasing the strength of the electromagnetic field, the equation of motion \eqref{lde_int_gen_y} turns into the ordinary Lorentz equation and its solutions cease to depend on $\bar{r}_e$. Another possibility is not to scale $x_\mu$ with $\bar{r}_e$ as in \eqref{subs_irregul}, but consider Eq. \eqref{lde_int_gen} with the solutions of the form $x^\mu(\tau/\bar{r}_e)$. Drawing the same reasonings, it is easy to see that these solutions are regular in $\bar{r}_e$ as well. As long as
\begin{equation}
    r_e=\frac{\bar{r}_e}{1+b\bar{r}_e/\e^{1/2}}=\frac{b^{-1}\e^{1/2}}{1+b^{-1}\e^{1/2}/\bar{r}_e},
\end{equation}
where $b$ is some constant, the regularity of a solution in $\bar{r}_e$ implies its regularity in the renormalized classical electron radius $r_e$ and the regularization parameter $\e^{1/2}$.

If we used another regularization of the Green function
\begin{equation}\label{green_func_reg}
    G^{-}_\e(x)=\frac{\theta(x^0)}{2\pi\e}\theta(x^2)g(x^2/\e),\qquad\int_0^\infty dxg(x)=1,
\end{equation}
then the asymptotics of the integral $I$ would become
\begin{equation}
    I\underset{\bar{r}_e\rightarrow0}{\rightarrow}4\e^{-3/2}\bar{r}_e^2\bar{y}_{[\mu}(s)\ups_{\nu]}\dot{y}^\nu(s)\int_0^\infty dtg'(t^2).
\end{equation}
The above arguments applied to this case reveal that there exists a regular limit $r_e\rightarrow0$ of solutions to the effective equations of motion of a charged particle with the regularization \eqref{green_func_reg} as well. A regularity of the electromagnetic field generated by this charged particle is also obvious. The regularization of the Green function is equivalent to the regularization of the current
\begin{equation}\label{current_reg}
    j^\mu(x)\;\rightarrow\;j^\mu_\e(x)=\Box_x\int dyG^+_\e(x-y)j^\mu(y),
\end{equation}
where $j_\mu(x)$ is the current of a point charge. The regularization $G^+_\e(x)$ of the advanced Green function is analogous to the retarded one \eqref{green_func_reg}. The effective equations of motion of a charge with the regularized current have the form \eqref{lde_int_gen}, but with the ``effective'' Green function (see for details, e.g., \cite{siss})
\begin{equation}
    G^{eff}_\e(x-y):=\int dzdz'\Box_x G^-_\e(x-z)G^-(z-z')\Box_{z'}G^+_\e(z'-y).
\end{equation}
It is not difficult to show that the effective regularized Green function is zero in the past light cone. A Poincare-invariance of this Green function implies
\begin{equation}\label{green_func_eff}
    G^{eff}_\e(x)=\frac{\theta(x^0)}{2\pi\e}\theta(x^2)\tilde{g}(x^2/\e)+\e^{-1}\theta(-x^2)h(-x^2/\e),
\end{equation}
where $\tilde{g}(x)$ is a generalized function satisfying the condition \eqref{green_func_reg}, while
\begin{equation}
    \int_0^\infty dxh(x)=0.
\end{equation}
The part of the effective Green function \eqref{green_func_eff} with the support lying outside of the light cone does not contribute to the integral in Eq. \eqref{lde_int_gen}. Thus, we revert to the case of the Green function regularization considered above. Notice also that we can use the retarded Green function of the form \eqref{green_func_reg_0} in the regularized current \eqref{current_reg} instead of the advanced Green function. It can be proven that the effective Green function takes the form \eqref{green_func_eff} in this case too.

When we pass from Eq. \eqref{lde_int_gen} to the LD equation \eqref{lde_ini}, we turn to a ``truncated'' description of a charge. Because of the truncation, spurious solutions arise which are non-regular in the coupling constants. They should be excluded, since only the regular solutions, which we call physical, to the LD equation are close to the solutions of \eqref{lde_int_gen} at small the regularization parameter $\e$.

Indeed, the LD equation is derived from \eqref{lde_int_gen} breaking off the series in $\e^{1/2}$. A regular in $r_e$ solution is regular in $\e^{1/2}$. If we substitute such a solution of the LD equation to Eq. \eqref{lde_int_gen} expanded in the series in $\e^{1/2}$, we shall ascertain that the obtained expression tends to zero with $\e\rightarrow0$ for any $\tau$. On the other hand, if we substitute a non-regular solution of the LD equation to Eq. \eqref{lde_int_gen} then the limit $\e\rightarrow0$ of the obtained expression may not exist. Making a general non-regular in $r_e$ ansatz to the LD equation \eqref{lde_ini}, it is not difficult to see that the non-regular solutions to LD equation have to possess an essentially singular point at $r_e=0$ provided some miraculous cancelations do not occur. Since it is the LD force which is responsible for the singularity, it should dominate over the Lorentz force at small $r_e$ and certain proper times $\tau$. Therefore, the non-regular solution to the LD equation tends to the non-regular solution \eqref{solution1d} to the free LD equation at $r_e\rightarrow0$. It can be directly verified from \eqref{lde_ini} stretching the proper time $\tau\rightarrow r_e\tau$. The solution \eqref{solution1d} does own the essentially singular point at $r_e=0$. But if we substitute this solution to Eq. \eqref{lde_int_gen}, it blows up at $\e\rightarrow0$ (the regularization parameter also enters the renormalized mass). So, the non-regular solutions cannot be regarded as approximate solutions to \eqref{lde_int_gen} at small the regularization parameter $\e$.

\begin{acknowledgments}

We appreciate Prof. V.G. Bagrov for stimulating discussions on the subject. The work is supported by the Russian Ministry of Education and Science, contract No 02.740.11.0238, the FTP ``Research and Pedagogical Cadre for Innovative Russia'', contracts No P1337, P2596, and the RFBR grant 09-02-00723-a.

\end{acknowledgments}

\end{document}